# Investigation on mass composition of UHE cosmic rays using CRPropa 2.0

**G Rastegarzade[1] B Parvizi[2]**
[1, 2,] Physics Department, Semnan University,
Semnan, P.O. Box 35196-45399, Iran

Email: G_ rastegar@alum.sharif.edu

**Abstract** In this paper, we used the Monte Carlo code CRPropa version 2.0 to simulate the propagation of UHE cosmic rays. In the simulation, uniform cosmic ray sources with spectral indices of $\alpha = 1.4$, $\alpha = 2$ and $E_{max} = 10^{21}$ eV for pure iron composition at the source position are assumed. We obtained the logarithmic average mass of cosmic rays as a function of energy and compared it with the experimental values in the ankle, and we demonstrated how the mass composition changes before and after the ankle.

## 1. Introduction

The sources and composition of ultra-high energy cosmic rays have not been specifically identified, and efforts continue to solve this mystery. The measurement of mass composition in high energy can help answer current questions about the origin of cosmic rays; in addition, the measurement provides a good clue for determining the causes of the knee, ankle, and GZK cut-off [1, 2] in the spectrum of the CRs.

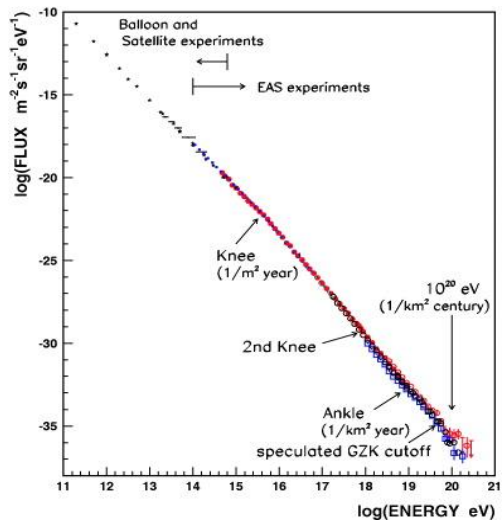

Fig. 1: Spectrum of cosmic rays

According to fig 1, the slope of the curve increases from 2.7 to 3.1 at approximately $10^{15}$ eV, and the number of particles decreases; this point is known as the knee. In addition, there is a second knee at $10^{17}$ eV that is not visible in this figure. At $10^{18}$ eV, the spectrum again experiences a sudden change: the slope decreases to 2.7, and the number of cosmic rays increases, which can be attributed to new sources in this energy and can be explained using the Larmor radius equation:

$$R_{larmor} = \frac{1}{Zec}\left(\frac{E_{CR}}{E_{eV}}\right)\left(\frac{\mu G}{B}\right) Kpc$$

This equation indicates that CRs with energies greater than $10^{18}$ eV cannot exist in the magnetic field of our galaxy. If the sources are in our galaxy, then the Larmor radius will be greater than the radius of our galaxy, and the CRs will be outside the galaxy.

Three models that have been under consideration regarding the subject of shifting from galactic to extra-galactic sources are as follows:

- Ankle model [3]: this model assumes that transfer occurs in the ankle, that the predominant composition is galaxy sources after the second knee, and that there are protons of extragalactic origin in the ankle.
- Dip model [4]: in this model, transfer occurs near the second knee, and there is a compositional change from galactic iron to extragalactic protons at this point; based on this model, there is an ankle because of the energy loss processes, such as pair production and pion production, as a result of proton interactions with the CMB.
- Mixed composition [5]: this model is similar to the dip model, except it considers the mixing of extragalactic particles instead of extragalactic protons in the place of transmission.

In this paper, we used the Monte Carlo code CRPropa version 2.0 [6, 7] to perform the simulations. The simulations can be performed either in one-dimensional (1D) or three-dimensional (3D) mode [8]. The deflections of UHECRs in extragalactic magnetic fields are considered in 3D mode. In 1D mode, cosmological and source evolution with redshift can be implemented.

## 2. Mass Composition of the CRs

The mass composition at ultra-low energies is the same as that of interstellar matter, which contains 90% hydrogen and 10% helium. Because the flux of cosmic rays decreases at high energies, it is much more complex and difficult to determine the mass composition at these energies. The Fly's Eye HiRes, Yakutsk, and Auger experiments operate in different energy ranges, and their mass composition results are associated with uncertainties at ultra-high energies. [9]

Fig. 8 shows the logarithmic mean mass of cosmic rays as a function of energy for $E > 10^{18}$ eV. We assumed a uniform source with $\alpha = 1.4$ and $E_{max} = 10^{21}$ eV for pure iron composition at the source position. The magnetic field for 3D simulation is equal to $B_x = B_y = 0$, $B_z = 10$ nG. Interactions of particles with the CMB are handled using the SOPHIA [10] event generator.

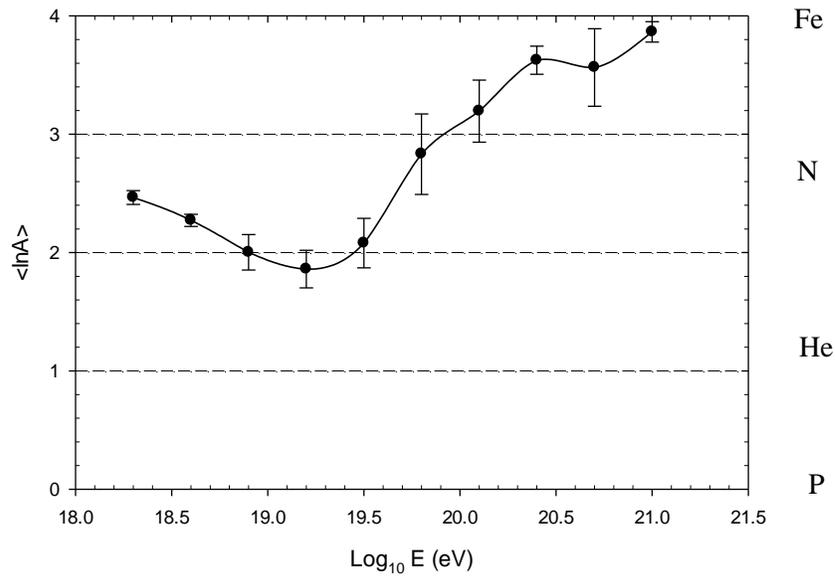

Fig. 2. Logarithmic mean mass of cosmic rays as a function of energy for pure iron composition, with $\alpha = 1.4$ and $E_{max} = 10^{21}$ eV from a 1D simulation. Points are from the average data of 5 cosmic rays.

As observed in fig. 2, the average mean mass decreases after $E = 10^{18}$ eV and increases after $10^{19}$ eV, which is consistent with the experimental results from Fly's Eye and Yakutsk experiments. Based on the Fly's Eye experiment, which uses $X_{max}$ for mass composition determination, the composition is heavy before $10^{18.5}$ eV and the spectrum is light above this energy [11]; this transition is called the "ankle". The mass composition spectrum changes when crossing the ankle. The Yakutsk experiment found a light spectrum above $10^{18.5}$ eV [12], and the HiRes experiment confirmed this result [13]. There is more iron at energies below the ankle energy, but the number of protons increases above the ankle energy. This experiment determined the location of the ankle to be at $10^{17.5}$ eV. The Auger experiment predicts a heavier spectrum at high energies compared with the results of the other experiments [14].

## 3. Influence of Magnetic Field on the Mass Composition of the Secondary CRs

Magnetic fields can play an important role in propagation and can increase the trajectory length of the particles during their journey from the source to the earth. Deflections of CRs due to the magnetic field are considered in the 3D simulation, but not in the 1D case. The redshift evolution of the background and the energy losses due to the adiabatic expansion of the universe are accounted for in the 1D simulation. To compare the mass of the secondary cosmic rays in the presence of a magnetic field, we obtained the average mass vs. energy for pure iron composition at the sources with α = 2 and 1.4. The results are shown in fig. 3.

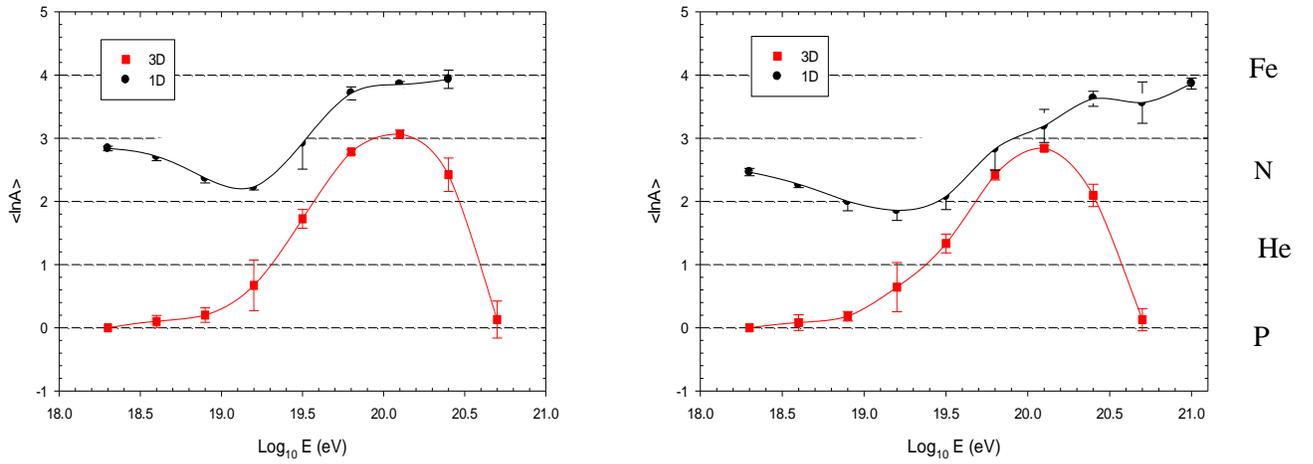

Fig. 3. Logarithmic mean mass of cosmic rays as a function of energy for pure iron composition, with α = 2 (left panel) and α = 1.4 (right panel) with $E_{max} = 10^{21}$ eV for a 1D simulation (circles) and a 3D simulation (squares). Points are from the average data of 5 cosmic rays.

The galactic magnetic field acts as a box for trapping charged particles and deflects them from their path, such that the particles will travel a longer path and undergo further interactions. As the number of interactions increases, the number of particles that decay into lighter particles also increases; as a result, the composition is lighter in the 3D simulation than it is in the 1D simulation.

## 4. Influence of the Spectral Indices on the Mass Composition of the CRs

Figure 4 shows a comparison of the mass composition for two different spectral indices: α = 2 and 1.4. For this comparison, the simulation was performed in both 1D and 3D mode for a source with iron composition.

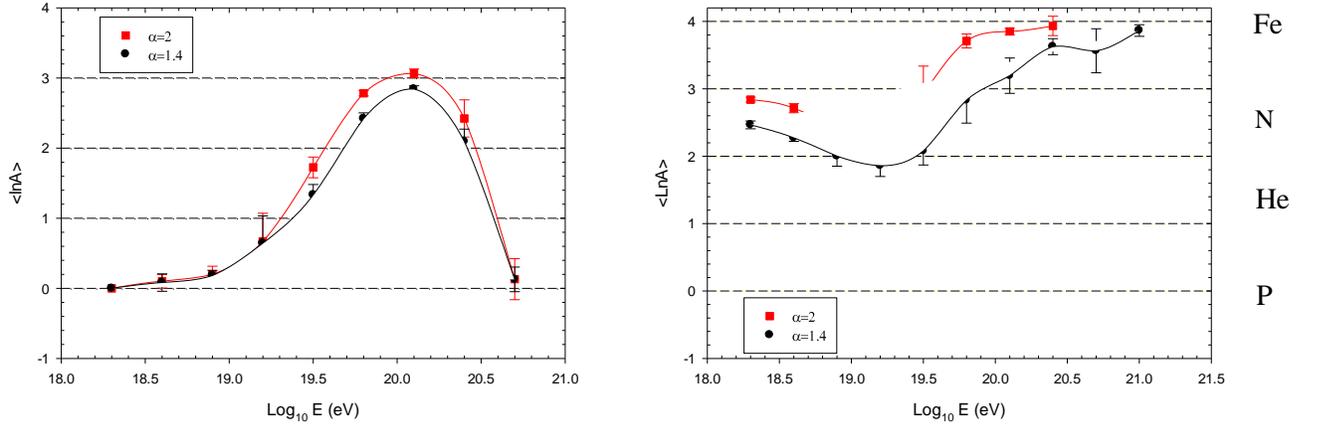

Fig. 4. Logarithmic mean mass of cosmic rays as a function of energy for pure iron composition, with α = 1.4 (circles) and α = 2 (squares) for $E_{max} = 10^{21}$ eV from a 3D simulation (left panel) and a 1D simulation (right panel). Points are from the average data of 5 cosmic rays.

As the spectral index increases (from α = 1.4 to α = 2), the number of particles decreases and the slope of the CR energy spectrum curve, which shows the flux as a function of energy, increases. Thus, a source with a smaller spectral index accelerates more particles to higher energies and is a strong source for the production of high-energy particles. Generally, the numbers of such sources are very small, which can be one of the reasons for the existence of an energy cutoff (GZK cut off) in the cosmic ray spectrum at energies above $6 \times 10^{19}$ eV. According to fig. 10, the mass composition is lighter when the spectral index is smaller. This behaviour occurs because particles with a smaller spectral index are more energetic than particles with a larger spectral index; thus, they interact more with cosmic microwave background (CMB) photons and the increment of interactions is equal to further decomposition of the nucleus and the conversion of the nucleus into smaller and lighter particles.

## 5. Conclusions

Using the Monte Carlo code CRPropa version 2.0, 5 uniform iron and proton sources with $E_{max} = 10^{21}$ eV and $\alpha = 1.4$ and 2 were simulated. For the 3D simulations, the magnetic field is set to 10 nG. We have determined the logarithmic average mass of cosmic rays as a function of energy. We have shown that after the ankle energy ($10^{18}$ eV), the mass composition is light and that after $10^{19}$ eV, it becomes heavy, which is consistent with the experimental results. In addition, we investigated the influence of the magnetic field on the mass composition and demonstrated that in the presence of the magnetic field, the mass composition is lighter because deflections caused by the magnetic field increase the propagation path length, thereby increasing the number of interactions, and the average mass number <A> decreases.

We found that the spectral index has an effect on the mass, i.e., as the spectral index increases (from α = 1.4 to α = 2), the mass composition becomes heavier.

## 6. References

[1] Greisen, K. 1966, Phys. Rev. Lett , 16, 748

[2] Zatsepin, G. T.  Kuzmin, V. A. 1966, Sov. Phys. JETP Lett, 4, 78

[3] Wibig, T. Wolfendale, A. W. 2004, Available from: <arXiv: 0410624v1>.

[4] Berezinsky, V. Grigorieva, S. 1988, *Astronomy & Astrophysics* 199, 1

[5] Aloisio, R. Berezinsky, V. S. Gazizov, A. 2012, *Astrop. Phys*, 39-40, 129

[6] Https:// crpropa. desy.de/images/0/05/User Guide.pdf

[7] https:// apcauger. in2p3. fr/CRPropa

[8] Sigl G,  Kulbartz J,  Maccioe L et al., 32$^{nd}$  ICRC (2011)

[9] Kampert, K. H. Unger, M. 2012, Astroparticle physics 35, 660-678

[10] Mucke, A. Engel, R. Rachen, J. P. Protheroe, R. J and Stanev, T. 2000, *Comput. Phys. Commun*, 124, 290

[11] Bird D. J., et al. 1993, (the HiRes Collaboration), *Phys. Rev. Lett*. 71, 3401.

[12] Knurenko, S. P. Sabourov, A. 2011, Nuclear Physics B (Proc. Suppl.) 212-213, 241-251

[13] Abbasi R. U., 2010, *Phys. Rev. Lett* , 104, 161101

[14] Abraham J. et al, 2010, (The Pierre Auger Collaboration), <arXiv: 1002.0699v1>